\documentclass[12pt]{article}
\usepackage{amsmath}

\newcommand{\bra}{\left\langle}
\newcommand{\ket}{\right\rangle}
\newcommand{\pder}[2]{\frac{\partial #1}{\partial  #2}}
\newcommand{\pdert}[2]{\frac{\partial^2 #1}{\partial  #2^2}}
\newcommand{\der}[2]{\frac{d #1}{d  #2}}

\newcommand{\vecr}{{\boldsymbol r}}
\newcommand{\vece}{{\boldsymbol e}}
\newcommand{\vecR}{{\boldsymbol R}}
\newcommand{\veceta}{{\boldsymbol \eta}}
\newcommand{\veck}{{\boldsymbol k}}

\newcommand{\ps}{p_{\rm s}}
\newcommand{\pst}{p_{\rm s}^{(2)}}
\newcommand{\Js}{J_{\rm s}}
\newcommand{\phis}{\phi_{\rm s}}
\newcommand{\qs}{q_{\rm s}}

\newcommand{\ve}{\varepsilon}
\newcommand{\e}{{\rm e}}
\newcommand{\mud}{\mu_{\rm d}}

\newcommand{\affiliation}[1]%
{\begin{center}{\small\it #1}\end{center}}
\renewcommand{\title}[1]{\begin{center}%
{\large\bf #1}\end{center}\par\bigskip}
\renewcommand{\author}[1]{\centerline{#1}}
\renewcommand{\maketitle}{}

\begin{document}

\title{ Long range spatial correlation between two Brownian particles 
under  external driving}

\author{Shin-ichi Sasa} 
\affiliation
{Department of Pure and Applied Sciences,
University of Tokyo, Komaba, Tokyo 153-8902, Japan}

\date{\today}

\begin{abstract} 
We study the large distance behavior of a steady distribution of two Brownian 
particles under external driving in a two-dimensional space. Employing 
a method of perturbative system reduction, we analyze a Fokker-Planck 
equation that describes the time evolution of the probability density for 
the two particles. The expression we obtain shows that there exists a 
long range correlation between the two particles, of $1/r^2$ type.
\end{abstract}

\maketitle


\section{Introduction}


The statistical properties of fluctuations at equilibrium are described 
by equilibrium statistical mechanics. This has been established
through  experimental measurements carried out to test the theoretical 
predictions of statistical mechanics.  In contrast to the equilibrium case, 
there is no known  general principle determining  the statistical properties 
of fluctuations under nonequilibrium conditions. Indeed, it might be thought 
that it is quite difficult to obtain  a universal theoretical framework 
on nonequilibrium fluctuations. 


There are many nonequilibrium steady states (NESSs) that settle down into 
an equilibrium state if one condition, such as the strength of an external 
driving force or the chemical potential at a boundary, is controlled. In 
such NESSs, the statistical properties of fluctuations can be elucidated
through an approach that seeks to determine how equilibrium fluctuations 
are modified under the influence of nonequilibrium conditions.


In a pioneering work in this context, Kuramoto studied open chemical systems
in 1974 and  pointed out that large scale fluctuations should be considered
separately from thermodynamic fluctuations occurring locally in space
\cite{ku74}. What is referred to in Ref. \cite{ku74} as 
``long range coherence'' is found more explicitly to appear  in the form 
of long range correlations of fluctuations for  conserved quantities 
\cite{LRC}.  Here,  there are two classes of
systems that exhibit long range correlation.  One class consists of 
systems driven by nonequilibrium boundary conditions. It includes 
laminar flow systems \cite{onuki78}, 
temperature gradient systems \cite{proc79} and 
density gradient systems  \cite{sp83}.
In such systems, anomalous  fluctuations originate from the spatial 
inhomogeneity of averaged quantities. The power law decay exponents
of the spatial correlation  for conserved quantities are basically 
of $1/r^{(d-2)}$ type, but different exponents can also be realized 
through composite effects \cite{ws}. 
The other type of systems exhibiting long range correlation are locally driven 
systems. In such systems, the statistical properties of local fluctuations 
differ substantially from those of  equilibrium systems. 
This modification yields  long range correlation of the form 
$1/r^d$ in $d$ $(\ge 2)$ dimensional 
systems \cite{ZWLV}. This behavior can be easily understood 
if we model the time development of a conserved quantity with
a phenomenological linear Langevin equation \cite{GLMS,Grin90}
in which the anisotropy of both the 
current noise intensity and the transportation coefficient without 
detailed balance is assumed.  


In this paper, we inquire whether long range correlation of $1/r^d$ 
type is peculiar to the fluctuations of macroscopic variables in driven 
systems. In general, a chain of correlated two-body interactions among  many 
particles provides a contribution to the correlation function for the 
density field. Thus, when we consider the system of a microscopic 
level (at which particle motion is described), it is reasonable 
to conjecture that long range correlation  appears only in the
macroscopic limit. However, if anisotropy with a local violation 
of detailed balance is the essence of long range correlation in driven 
systems, it may not be necessary to have a many body system
in order to observe such correlation. As an extreme case, 
a system consisting of two particles under external driving 
may exhibit long range correlation.


With this motivation, quite recently, in a  calculation of the steady 
probability for the positions of two interacting random walkers in a 
$d$ $(\ge 2)$ dimensional lattice under external driving, 
it has been found that the large distance behavior of the 
probability, including the existence or non-existence of 
long range correlation, depends on the choice of the transition rules 
satisfying the condition of 
local detailed balance \cite{hal04}. 
It is surprising that there is such a dependence, 
considering the fact that universal  relations 
in the linear response regime do not depend on the precise nature of
these rules. Given this situation, we are led to ask, 
Which rule is physically meaningful? However, it is difficult 
to answer this question by considering such transition rules themselves.
For this reason, we investigate a physical model that  
corresponds directly to an experimental system. 


In the present paper, we study the large distance behavior 
of two Brownian particles with a local interaction under nonequilibrium 
conditions. Specifically, we consider two Brownian particles in a two 
dimensional space of temperature $T$. Their interaction is characterized 
by an interaction length $\xi$. They are driven in one direction, which 
we choose as the $x$ direction, 
by a constant external force $f$, under the influence 
of a periodic potential with period $\ell$. Let $\pst(\vecr_1,\vecr_2)$ be 
the steady probability density for the positions of the  
two particles, $(\vecr_1,\vecr_2)$.  Calculating 
$\pst(\vecr_1,\vecr_2)$ to leading order of  in the asymptotic limit 
$|\vecr_1-\vecr_2| \to \infty$ under the assumption that $\xi \gg \ell$,
we  find that 
\begin{equation}
\pst(\vecr_1,\vecr_2)-  \ps(\vecr_1) \ps(\vecr_2)
\simeq  \frac{c(\vecr_1,\vecr_2)}{|\vecr_1-\vecr_2|^2},
\end{equation}
where  $c(\vecr_1,\vecr_2)$ does not depend on $|\vecr_1-\vecr_2|$, but
on the direction of the vector $\vecr_1-\vecr_2$, and $\ps(\vecr)$ is the one
body steady probability density for a system without an interaction.
Thus, it is found that a long range correlation of  $1/r^d$ type  
appears in  a system consisting of two locally 
interacting particles under external driving.


In order to obtain this result, we  apply a method of perturbative 
system reduction to the Fokker-Planck equation describing the time 
development of the probability density in the system described above.
Here, what we refer to as ``perturbative system reduction'' consists
of a perturbative calculation designed to obtain a simpler representation
of a dynamical 
system by restricting what we wish to describe.
The first application of perturbative system reduction to reaction 
diffusion systems was carried out by Kuramoto and Tsuzuki \cite{kuracgl}.
They obtained a complex Ginzburg-Landau equation near a Hopf bifurcation  
and subsequently derived the simplest partial differential equation 
exhibiting  spatially extended chaos \cite{kuraKS}, which is now 
called the Kuramoto-Sivashinsky equation \cite{kurabook}.  
These methods of derivation have matured since that time,
and the universal structure underlying the calculations 
has been elucidated \cite{kura84,kura89}. 


This paper is organized as follows. In Section 2, we present the
stochastic model we study and describe the basic features of the model. 
In Section 3, we formulate a perturbative system reduction by first 
reviewing  the basic ideas  introduced by Kuramoto.  In Section 4, 
performing a perturbative expansion of the model, we obtain a reduced 
model describing the large scale behavior of the system in question.
Using this result, in Section 5, we derive the asymptotic form 
of the large distance behavior of the steady probability density.
Section 6 is devoted to concluding remarks.
 
\section{Model}

We study the motion of two small particles (on the oder of micro-meters 
in radius) interacting with each other in a fluid of temperature $T$. 
The particles are confined to a two dimensional square of length $L$  
and are subject to a periodic potential $U$ of period 
$\ell$ in a single direction, which we chose as the $x$ direction.  
Typical systems with such properties can be realized experimentally
\cite{hara,KTG}. 
Further, a flow with constant velocity can be used to apply a constant 
driving force $f$ to the particles in the $x$ direction.  
In this way, it is possible to experimentally realize NESSs
for such a particle system.

Let $(\vecr_1,\vecr_2)$ represent the  positions of the particles. We assume
that their motion is described by the Langevin equation  
\begin{equation}
\gamma\dot{\vecr_i}=f-\frac{\partial  U(\vecr_i)}{\partial \vecr_i}
-\frac{\partial V(\vecr_1-\vecr_2)}{\partial \vecr_i}
+\sqrt{2\gamma T}\veceta_i(t),
\label{lan} 
\end{equation}
where $\vecr=(x,y)$, $V(\vecr_1-\vecr_2)$ is an interaction potential 
with interaction length $\xi$, (e.g. $V(\vecr)=\ $ const. for  $|\vecr| 
\ge \xi.)$, and $U(\vecr)$ is a periodic potential satisfying
\begin{equation}
U(\vecr+\ell \vece_x)=U(\vecr).
\end{equation}
Further, $\veceta_i(t)$ represents Gaussian white noise with zero mean 
and unit dispersion.  Here, the Boltzmann constant is set to unity.  
For simplicity, we assume periodic boundary conditions in both 
directions and that the system size $L$ is sufficiently larger than
$\xi$ and $\ell$. 

In this system, the probability density of the particle positions,
$p( \vecr_1,\vecr_2,t) $, obeys the Fokker-Planck equation 
\begin{equation}
\pder{p}{t}=\frac{1}{\gamma}\sum_{i=1}^2
\pder{}{\vecr_i}
\left[-f\vece_x p+
\frac{\partial  U(\vecr_i)}{\partial \vecr_i}p
+\frac{\partial V(\vecr_1-\vecr_2)}{\partial \vecr_i}p
+T\pder{p}{\vecr_i} 
\right].
\label{fp}
\end{equation}
Although we wish to obtain the steady solution of (\ref{fp}),
it seems unfeasible to derive such a solution in exact form.
For this reason, we formulate a perturbation method to extract the 
large scale behavior of the steady state solution under some assumptions. 
Before presenting the analysis, some preparation is needed. 

We first note that when $V(\vecr_1-\vecr_2)=0$, 
the steady state solution of (\ref{fp}) can be  derived easily
in the form $\ps(\vecr_1;f)\ps(\vecr_2;f)$, where 
\begin{equation}
\ps(x,y;f)=\frac{1}{Z} I_-(x),
\label{ps}
\end{equation}
with
\begin{equation}
I_-(x)=\int_0^\ell dx' \e^{-\beta U(x)+\beta U(x'+x)-\beta fx'}.
\end{equation}
Here,  $\beta=1/T$ and  $Z$ is a normalization factor that is chosen
so that we have 
\begin{equation}
\int_0^{\ell}dx\ps(x,y;f)=\ell.
\end{equation}
For later convenience, we define the following quantities:
\begin{eqnarray}
\Js(f) &=&\frac{1}{\gamma}\left( f-\pder{U(\vecr)}{x} \right)
\ps(\vecr;f)-\frac{T}{\gamma}
\pder{\ps(\vecr;f)}{x}, \label{jdef} \\
\phis(\vecr;f) &=& \log \ps(\vecr;f). \label{pdef}
\end{eqnarray}
Note that $\Js(f)$ does not depend on $\vecr$, because  
$\Js(f)$ corresponds to the one particle steady state 
probability current for the case $V=0$. 

\section{Formulation}

Next, we consider the effect of the interaction potential $V$ 
by first writing
\begin{equation}
p(\vecr_1,\vecr_2,t) = \ps(\vecr_1;f_1)\ps(\vecr_2;f_2)q(\vecr_1,\vecr_2,t),
\label{qdef}
\end{equation}
where $f_1$ and $f_2$ are configuration dependent forces defined by
\begin{equation}
f_i(\vecr_1,\vecr_2)=f-\pder{V(\vecr_1-\vecr_2)}{x_{i}}.
\end{equation}
Then, substituting (\ref{qdef}) into (\ref{fp}), we obtain the evolution 
equation of $q(\vecr_1,\vecr_2,t)$ as
\begin{equation}
\pder{q}{t}=\sum_{i=1}^2 
\left[ \hat M_i q -\frac{1}{\ps(\vecr_i;f_i)}
\der{\Js(f_i)}{f_i}\pder{f_i}{x_i}q
+\frac{1}{\gamma} \frac{\partial}{\partial y_i}
\left( \pder{V}{y_i} q  \right)
\right],
\label{qevol}
\end{equation}
where the operator $\hat M_i$ is defined as
\begin{equation}
\hat M_i\equiv  -\frac{ \Js(f_i)}{\ps(\vecr_i;f_i)}\pder{}{x_i}
+\frac{T}{\gamma}\pder{\phis(\vecr_i;f_i)}{x_i}\pder{}{x_i}
+\frac{T}{\gamma}\left( 
\frac{\partial^2}{\partial x_i^2}+
\frac{\partial^2}{\partial y_i^2} \right).
\end{equation}

We study a simple situation where we fix  the period $\ell$ of 
the periodic potential $U(\vecr)$ and make the range $\xi$ of 
the interaction longer and longer. 
We then introduce a small parameter $\ve$ representing 
the extent of the scale separation: $\ve \equiv \ell/\xi \ll 1$. 
The existence of two typical length scales is account for  explicitly 
by introducing the large scale coordinate $\vecR_i =(X_i,Y_i)
\equiv \ve \vecr_i$ and the  periodic coordinate $\theta_i\equiv 
{\rm mod}(x_i,\ell)$, which expresses the $i$-th particle position 
in terms of the phase of the periodic potential. 
Using these coordinates, we rewrite $U$ and $V$ as 
\begin{eqnarray}
 U(\vecr_i) &=& \tilde U(\theta_i)  \\
V(\vecr_1-\vecr_2) &=& \tilde V(\vecR_1-\vecR_2).
\end{eqnarray}
In a similar way, we further define $\tilde \ps(\theta_i;\tilde f_i)$
and $\tilde \phis(\theta_i;\tilde f_i)$, where 
\begin{equation}
\tilde f_i(\vecR_1,\vecR_2) 
= f-\ve \pder{\tilde V(\vecR_1-\vecR_2)}{X_i}.
\end{equation}

Now, we introduce a slowly varying field $Q(\vecR_1,\vecR_2,t)$
in such a way that 
\begin{equation}
q(\vecr_1,\vecr_2,t)=  Q(\vecR_1,\vecR_2,t)+
\ve \rho^{(1)}(\theta_1,\theta_2; [Q])+
\ve^2 \rho^{(2)}(\theta_1,\theta_2; [Q]) +\cdots,
\label{qQ}
\end{equation}
where $g([Q])$ represents the functional dependence of $g$ on 
$Q(\vecR_1,\vecR_2,t)$. At this stage, $Q$ is not determined.
According to Ref. \cite{kura89} written by Kuramoto in 1989, 
$Q$ can be regarded as the coordinate of a point on the slow 
manifold in the functional space $\{q \}$. With this interpretation,
(\ref{qQ}) provides a representation of the slow manifold
in terms of $Q$. 
Here, obviously, such a representation can be chosen 
arbitrarily.  We therefore choose the following convenient
form of the time evolution of $Q$ is expressed by
\begin{equation}
\pder{Q}{t}=\ve \Omega^{(1)}([Q])+\ve^2 \Omega^{(2)}([Q])+\cdots.
\label{Qevol}
\end{equation}
If $Q$ can be determined uniquely with this requirement, then 
(\ref{Qevol}) represents the system reduction we seek.
Equations (\ref{qQ}) and (\ref{Qevol}) constitute the basic assumptions 
of the perturbative system reduction for (\ref{qevol}), with the 
replacements $U$ by $\tilde U$, and so on. Thus, the problem we face is
to determine whether $\rho^{(n)}(\theta_1,\theta_2; [Q])$ and 
$\Omega^{(n)}([Q])$ ($n=1,2,\cdots$) can be determined in an essentially 
unique way.  In the next section, we  see that indeed this can be done.

\section{Analysis}

We first  substitute (\ref{qQ}) and (\ref{Qevol}) into (\ref{qevol})
and extract all  terms proportional to $\ve$. We then obtain 
\begin{equation}
\Omega^{(1)}([Q])= \sum_{i=1}^2 
[\hat M_i^{(0)} \rho^{(1)}+\hat M_i^{(1)} Q],
\label{1st}
\end{equation}
where the operators $\hat M_i^{(0)}$ and $\hat M_i^{(1)}$ are given by
\begin{eqnarray}
\hat M_i^{(0)} &\equiv &  - \frac{\Js(\tilde f_i)}{ \tilde \ps(\theta_i;\tilde f_i)}
\pder{}{\theta_i}
+\frac{T}{\gamma}
\pder{\tilde\phis(\theta_i;\tilde f_i)}{\theta_i}\pder{}{\theta_i}
+\frac{T}{\gamma} \frac{\partial^2}{\partial \theta_i^2}, \\
\hat M_i^{(1)} &\equiv&  
-\frac{\Js(\tilde f_i)}{ \tilde \ps(\theta_i;\tilde f_i)}
\pder{}{X_i}
+\frac{T}{\gamma}
\pder{\tilde \phis(\theta_i;\tilde f_i)}{\theta_i}\pder{}{X_i}
+2\frac{T}{\gamma} \frac{\partial^2}{\partial \theta_i \partial X_i}.
\end{eqnarray}
Because $\hat M_i^{(0)} \cdot 1=0$,  $\rho^{(1)}$ can be obtained
only when the solvability condition is satisfied. Thus, we need 
to  find an explicit form of the solvability condition.
 
Let us define an operator $\hat L_i$  as
\begin{equation}
\hat L_i \cdot \equiv \frac{1}{\gamma} \pder{}{\theta_i} 
\left( \pder{\tilde U}{\theta_i} \cdot +T \pder{}{\theta_i} \cdot \right).
\end{equation}
Then, using the relation
\begin{equation}
\hat L_i (\tilde \ps(\theta_i) \phi(\theta_i))=
\tilde \ps(\theta_i) (\hat M_i^{(0)}\phi(\theta_i)),
\end{equation}
for an arbitrary square integrable periodic function $\phi(\theta_i)$,
we obtain
\begin{equation}
\int d\theta_i \tilde \ps(\theta_i) (\hat M_i^{(0)} \phi(\theta_i))=0.
\end{equation}
{}From this, the solvability condition for (\ref{1st})  turns out to be
\begin{equation}
\int d\theta_1d\theta_2 \tilde \ps(\theta_1)\tilde \ps(\theta_2)
\left[ \Omega^{(1)}([Q])- \sum_{i=1}^2 \hat M_i^{(1)} Q \right]=0,
\label{1st:sc}
\end{equation}
which yields
\begin{equation}
\Omega^{(1)}([Q]) = - \sum_{i=1}^2 \Js(\tilde f_i) \pder{Q}{X_i}.
\label{1st:sc2}
\end{equation}
Under this condition, we can derive $\rho^{(1)}$ of  (\ref{1st})
in the form 
\begin{equation}
\rho^{(1)}= \sum_{i=1}^2  a(\theta_i;f_i)
 \pder{Q}{X_i}+\chi^{(1)}(\vecR_1,\vecR_2 ),
\end{equation}
where $a(\theta_i;f_i)$ can be calculated explicitly as in 
Ref. \cite{HSIII}, and $\chi^{(1)}(\vecR_1,\vecR_2)$ is an arbitrary 
function of $(\vecR_1,\vecR_2)$.
(The choice of this function does not influence the result for 
$\Omega^{(n)}$.)

Next, we sum up all terms proportional to $\ve^2$. This yields
\begin{eqnarray}
\Omega^{(2)}([Q])+\frac{\delta \rho^{(1)}}{\delta Q}\cdot \Omega^{(1)}([Q])
&=& \sum_{i=1}^2 
\left[
\hat M_i^{(0)} \rho^{(2)}+\hat M_i^{(1)} \rho^{(1)}
+\hat M_i^{(2)} Q  \right.\nonumber \\
& +& \left.  \frac{\mud(\tilde f_i)}{\tilde \ps(\theta_i;\tilde f_i)}
\pdert{\tilde V}{X_i}Q
+\frac{1}{\gamma} \frac{\partial}{\partial Y_i}
\left( \pder{\tilde V}{Y_i} Q  \right)
\right],
\label{2nd}
\end{eqnarray}
where
\begin{equation}
\hat M_i^{(2)} = \frac{T}{\gamma} \left( 
\frac{\partial^2}{\partial X_i^2}+
\frac{\partial^2}{\partial Y_i^2} \right),
\end{equation}
and we have introduced the differential mobility $\mud(f)\equiv 
d \Js/df$,  which is found to be 
\begin{equation}
\mud(f)= \frac{1}{\gamma} 
\frac{\frac{1}{\ell}\int_0^\ell dx I_-(x)I_+(x)}
{\left(\frac{1}{\ell}\int_0^\ell dx I_-(x) \right)^2},
\label{dmob}
\end{equation}
with
\begin{equation}
I_+(x)=\int_0^\ell dx' e^{\beta U(x)-\beta U(x-x')-\beta fx'}.
\end{equation}
The solvability condition for $\rho^{(2)}$ in (\ref{2nd}), which
is obtained by multiplying both sides of (\ref{2nd}) by
$\int d\theta_1d\theta_2 \tilde 
\ps(\theta_1)\tilde \ps(\theta_2)$, yields
\begin{equation}
\Omega^{(2)}=\sum_{i=1}^2
\left[ \mud(\tilde f_i) \pdert{\tilde V}{X_i}Q
      +D(f)\pdert{Q}{X_i} 
+\frac{T}{\gamma}  \pdert{Q}{Y_i} 
+\frac{1}{\gamma}  \frac{\partial}{\partial Y_i}
\left( \pder{\tilde V}{Y_i} Q \right)
\right] ,
\label{sc2}
\end{equation}
where $D$ is obtained  as 
\begin{equation}
D(f)=\frac{T}{\gamma} 
\frac{\frac{1}{\ell}\int_0^\ell dx (I_-(x))^2 I_+(x)}
{\left(\frac{1}{\ell}\int_0^\ell dx I_-(x) \right)^3}.
\label{dif}
\end{equation}
This quantity clearly represents the diffusion constant 
in the $x$ direction \cite{RH}. 
The derivation here is parallel to that given in Ref. \cite{HSIII}.
Under the solvability condition (\ref{sc2}), we can obtain
$\rho^{(2)}$.   

Finally, carrying out a similar calculation, we can determine 
$\rho^{(n)}(\theta_1,\theta_2; [Q])$ and $\Omega^{(n)}([Q])$ 
from the terms proportional to $\ve^n$ in (\ref{qevol}), with
(\ref{qQ}) and (\ref{Qevol}). This iterative procedure constitutes
the perturbative system reduction.

\section{Long range correlation}

Recall that our  goal is to obtain the steady state solution of the 
Fokker-Planck equation (\ref{fp}) under the assumption $\ve \ll 1$. 
Let us express this solution as 
\begin{equation}
\pst(\vecr_1,\vecr_2)=\ps(\vecr_1) \ps(\vecr_2) \qs(\vecr_1,\vecr_2).
\end{equation}
Then, as far as we focus on the large distance behavior of 
$\qs(\vecr_1,\vecr_2)$,  from (\ref{qdef}), (\ref{qQ}), (\ref{Qevol}),
(\ref{1st:sc2}) and (\ref{sc2}), 
we can assume that $\qs(\vecr_1,\vecr_2)$ satisfies the
equation
\begin{equation}
\sum_{i=1}^2 
\left[- \Js(f)\pder{\qs}{x_i} 
      + \mud(f) \frac{\partial}{\partial x_i}
        \left( \pder{V}{x_i} \qs \right)
      +D(f)  \pdert{\qs}{x_i}
+\frac{T}{\gamma} \pdert{\qs}{y_i} 
+\frac{1}{\gamma}  \frac{\partial}{\partial y_i}
\left( \pder{V}{y_i} \qs \right) \right]=0.
\label{qs}
\end{equation}

Now, we define $\psi(\vecr_1,\vecr_2)$, which represents 
the non-equilibrium contribution of $\qs$, through the relation
\begin{equation}
\qs(\vecr_1,\vecr_2)
=\e^{-\beta V(\vecr_1-\vecr_2)+\psi(\vecr_1,\vecr_2)}.
\end{equation}
Furthermore, assuming that $V$ is sufficiently small,
we linearize (\ref{qs}) with respect to $\psi$ 
and $V$. This yields 
\begin{equation}
\sum_{i=1}^2 
\left[
\left\{ - \Js(f) +(D(f) - \mud(f) T) \pder{}{x_i} \right\}
\left( -\beta \pder{V}{x_i}+\pder{\psi}{x_i}  \right)
+\frac{T}{\gamma} \left(\pdert{}{x_i} +\pdert{}{y_i}\right)\psi 
\right]
=0.
\label{psieq}
\end{equation}
Then, using the Fourier expansion, we can solve for $\psi$ in 
(\ref{psieq}) as
\begin{equation}
\psi(\vecr_1,\vecr_2)= (D-\mud T)\beta 
\int \frac{d^2 \veck}{(2\pi)^2} \e^{-i\veck (\vecr_1-\vecr_2)}
\frac{k_x^2}{D k_x^2+ T k_y^2/\gamma} \hat V(\veck),
\end{equation}
where 
\begin{equation}
\hat V(\veck)=
\int d^2 \vecr \e^{i\veck \vecr}V(\vecr).
\end{equation}
{}From this result, it is straightforward to derive 
the asymptotic form 
\begin{equation}
\psi(\vecr_1,\vecr_2) \simeq
\frac{1}{2\pi}
\frac{D-\mud T}{(DT)^{3/2}}\gamma^{1/2}\hat V(0)
\frac{(x_1-x_2)^2/D-(y_1-y_2)^2\gamma/T}
{((x_1-x_2)^2/D+(y_1-y_2)^2\gamma/T)^2}
\end{equation}
in the  region  $ \xi \ll |\vecr_1-\vecr_2| \ll L$.
As shown in Ref. \cite{HSIII}, the Einstein relation $D=\mud T$ 
is violated in general NESSs. Thus, we conclude that there is  
long range correlation  of  $1/r^2$ 
type.

\section{Concluding remarks}

We have demonstrated that there exists long-range spatial correlation
between  two interacting Brownian particles under  external driving.
We have found that this long range correlation is proportional to 
$D-\mud T$, which represents the degree of the breakdown of  detailed 
balance. It is quite reasonable to expect that the long range correlation
found for this two particle system exists also in many particle
systems, with a quantitative correction arising from  many body effects.
It is a future project to study  many body effects by extending approach
in the present paper.

The existence of long range correlation of  $1/r^d$  type makes it difficult
to construct a universal framework for a statistical theory. 
Let us explain the reason for this by considering 
$N$ particles in a two dimensional box of  length $L$. 
We write the $N$-body steady state distribution for this system as 
\begin{equation}
p(\{\vecr_i\})=\e^{-\beta \Phi(\{\vecr_i \})} .
\end{equation}
One may naively interpret $\Phi (\{\vecr_i \})$  as 
``effective energy `` of the particles under nonequilibrium conditions,
because $\Phi (\{\vecr_i \})$ corresponds to the Hamiltonian of 
the particle configuration $\{ \vecr_i \}$ at equilibrium.
With this interpretation, there appears an effective 
long range interaction potential of $1/r^2$ type.
Then, for such systems, simple considerations yield the estimation 
\begin{equation}
\bra \Phi \ket \sim  L^2 \log L
\label{nonex}
\end{equation}
in the limit that $L \to\infty$, with fixed $L^2/N$. This implies that
extensivity, which is the most essential property of thermodynamic systems, 
does not hold. Such statistical systems are pathological for the following 
reasons. First, a statistical distribution located in the central region 
of the system  depends sensitively on the nature of the  boundary
conditions \cite{Eyink98}. That is, it is difficult to define a  bulk
region for the system.  Second, if $\Phi$ can be measured as 
``energy'' using some experimental method, (\ref{nonex}) implies that 
a significant amount of energy can be extracted by merely splitting one 
system or combining two systems. 
Because such a situation seems to be unphysical, 
it is reasonable to conjecture that $\Phi$ does not represent
an ``energy''.  These conclusions  cast doubt on the possibility of
realizing a unified statistical framework. 

Despite these seemingly intractable properties, we  wish to seek 
a universal statistical framework for NESS by separating the problem in the 
following way. First, we propose to check the possibility that large scale 
fluctuations can be distinguished from small scale fluctuations in some way.
If this can be done,  we  hope to determine whether small scale fluctuations, 
which may deviate substantially from those of an equilibrium system, 
can be characterized in terms of an energetic quantity. Then, finally, 
we hope to study large scale fluctuations on the basis of the  
characterization of smalls scale fluctuations.  
Recently, we have made some progress in the characterization of small scale 
fluctuations through an extension of thermodynamic  
functions \cite{Sa-Ta,HSI}.  We now propose to attempt unifying  large 
scale anomalous  fluctuations with our thermodynamic
framework, going  beyond the result of  the present study.

In closing this paper, we would like to return to  
Ref. \cite{ku74} written by Kuramoto in 1974.  The observation made
there that large scale fluctuations should 
be considered separately led Kuramoto  to focus on dynamical 
behavior of macroscopic variables. In particular,  when  solutions 
of deterministic equations for macroscopic variables describe a 
rich variety of phenomena including oscillations and chaos, the 
understanding of such phenomena from a dynamical system point of view
may be most important.  With this realization, Kuramoto naturally was led
to study dynamical systems.  
This is  regarded as the  genesis of nonlinear dynamics  
as a method for studying nonequilibrium statistical phenomena.
The most important  message here seems to be  that to formulate 
questions that do not conform to the contemporary mainstream 
can lead to  new fields of research. Today, the study of nonlinear 
dynamics has been fully developed. Following Kuramoto, 
we should consider to seek the formulation of precise and deep questions 
regarding nonlinear and nonequilibrium systems that do not conform to
current trends.

\section*{Acknowledgments}

It is a great pleasure to dedicate this paper to Prof. Kuramoto on the
occasion of his retirement from Kyoto University. I have learned the nature
of ``true study'' as well as many scientific ideas  from him since the time
I was a graduate student under his supervision. I also thank 
H. Tasaki for extensive discussions on statistical mechanics in NESS.


\end{document}